\documentclass[aps,pra,twocolumn,superscriptaddress]{revtex4}
\usepackage{graphicx}
\usepackage{amssymb}
\usepackage{amsmath}
\usepackage{bm}
\begin{document}

\title[]{Observation of vortex-antivortex pairing in decaying 2D turbulence of a superfluid gas}

\author{Sang Won Seo}
\affiliation{Department of Physics and Astronomy, and Institute of Applied Physics, Seoul National University, Seoul 08826, Korea}
\affiliation{Center for Correlated Electron Systems, Institute for Basic Science, Seoul 08826, Korea}

\author{Bumsuk Ko}
\affiliation{Department of Physics and Astronomy, and Institute of Applied Physics, Seoul National University, Seoul 08826, Korea}
\affiliation{Center for Correlated Electron Systems, Institute for Basic Science, Seoul 08826, Korea}

\author{Joon Hyun Kim}
\affiliation{Department of Physics and Astronomy, and Institute of Applied Physics, Seoul National University, Seoul 08826, Korea}

\author{Y. Shin}\email{yishin@snu.ac.kr}
\affiliation{Department of Physics and Astronomy, and Institute of Applied Physics, Seoul National University, Seoul 08826, Korea}
\affiliation{Center for Correlated Electron Systems, Institute for Basic Science, Seoul 08826, Korea}

\begin{abstract}
In a two-dimensional (2D) classical fluid, a large-scale flow structure emerges out of turbulence, which is known as the inverse energy cascade where energy flows from small to large length scales. An interesting question is whether this phenomenon can occur in a superfluid, which is inviscid and irrotational by nature. Atomic Bose-Einstein condensates (BECs) of highly oblate geometry provide an experimental venue for studying 2D superfluid turbulence, but their full investigation has been hindered due to a lack of the circulation sign information of individual quantum vortices in a turbulent sample. Here, we demonstrate a vortex sign detection method by using Bragg scattering, and we investigate decaying turbulence in a highly oblate BEC at low temperatures, with our lowest being $\sim 0.5 T_c$, where $T_c$ is the superfluid critical temperature. We observe that weak spatial pairing between vortices and antivortices develops in the turbulent BEC, which corresponds to the vortex-dipole gas regime predicted for high dissipation. Our results provide a direct quantitative marker for the survey of various 2D turbulence regimes in the BEC system.
\end{abstract}

\maketitle

\section{Introduction}

Quantum turbulence (QT) is a state of chaotic flow in a superfluid. Because of its inviscidity and quantized circulation, QT constitutes a unique realm in turbulence research. Decades of study involving superfluid helium have revealed many aspects of QT similar to and different from those of turbulence in classical fluids~\cite{Skrbek2012,Tsubota2013}, and atomic Bose-Einstein condensates (BECs) were recently employed to extend the scope of QT studies~\cite{Henn2009,Neely2013,Kwon2014,Navon2016}. One of the experimentally unanswered questions is related to the inverse energy cascade in two-dimensional (2D) QT. It is well known that regarding the 2D turbulence of a classical hydrodynamic fluid, the kinetic energy flows toward large length scales, generating a large-scale flow structure due to small-scale forcing~\cite{Krichnan1980}. This phenomenon is qualitatively different from three-dimensional turbulence, where energy is dissipated at small length scales. The key issue regarding 2D QT is whether the inverse energy cascade occurs and consequently leads to the formation of a large superflow structure; this issue has drawn a great deal of recent theoretical attention~\cite{Horng2009,Numasato2010,White2012,Bradley2012,Nowak2012,Kusumura2013,Reeves2013,Chesler2013,Simula2014,
Reeves2014,Billam2014,Billam2015,Stagg2015,Du2015,Skaugen2016}. Two-dimensional QT is also relevant to the 2D superfluid phase transition which is associated with free vortex proliferation in the Berezinskii-Kosterlitz-Thouless description~\cite{Nazarenko14}.

\begin{figure}[b]
\centering
\includegraphics[width=7.3cm]{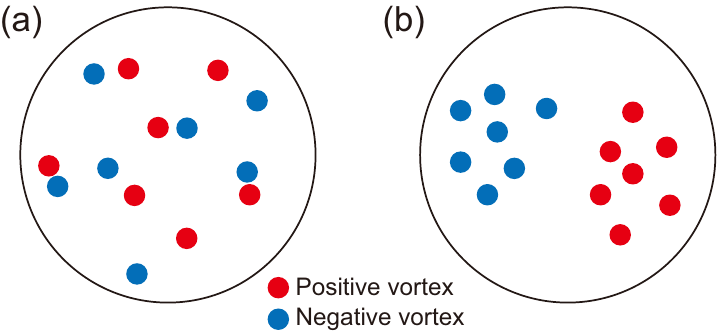}
\caption{Two vortex configurations of neutral 2D quantum turbulence. (a) Each vortex has an opposite-sign vortex as its nearest neighbor, and the mean kinetic energy per vortex, $\varepsilon_v$, of the system is low. (b) Vortices with the same circulation signs are clustered, and a large vortex dipole structure is formed in the system, having high $\varepsilon_v$. This is the Onsager vortex state expected from the inverse energy cascade in 2D.}
\label{fig1}
\end{figure}

The turbulent flow of an irrotational superfluid is characterized by the configuration of quantum vortices in the superfluid. In 2D, quantum vortices are topological point defects, and the turbulent superfluid can be depicted as a system of interacting `vortex' particles. This point-vortex picture was introduced by Onsager in his model, which presented a statistical description of classical 2D turbulence~\cite{Onsager1949,Eyink2006}. The turbulent state is parameterized with the mean vortex energy, $\varepsilon_v=E_v/N_v$, where $E_v$ is the incompressible kinetic energy of the system and $N_v$ is the total vortex number~\cite{Yatsuyanagi2005,Simula2014}. Figure 1 illustrates two vortex configurations for low and high $\varepsilon_v$ values. In Fig.~\ref{fig1}(a), each vortex is adjoined by an antivortex, i.e., a vortex with opposite circulation and their velocity fields cancel each other out in the far-field, thus lowering $\varepsilon_v$. A small dipole of the vortex and antivortex undergoes a linear motion, and the low-$\varepsilon_v$ states are referred to as a vortex-dipole gas regime~\cite{Reeves2014,Simula2014,Billam2015}. On the other hand, Fig.~\ref{fig1}(b) shows a high-$\varepsilon_v$ state, where vortices of same circulation signs are clustered, constructively enhancing the superflow velocity. Large vortex clusters are called Onsager vortices, which are anticipated to develop as a result of the inverse energy cascade~\cite{Reeves2013,Reeves2014,Billam2014,Simula2014,Billam2015}.

How the mean vortex energy $\varepsilon_v$ changes as the vortex system evolves underlies the inverse energy cascade problem in 2D QT. The system eventually evolves toward a stationary ground state by decreasing both $E_v$ and $N_v$ via various dissipation mechanisms, such as sound radiation~\cite{Demircan1996}, mutual friction by coexisting thermal components~\cite{Kobayashi2006,Berloff2007,Moon2015}, and vortex-antivortex pair annihilation~\cite{Berloff2007,Kwon2014}. It has been noted that the vortex-antivortex annihilation would facilitate the increase of $\varepsilon_v$ because the contribution of the annihilated vortex dipole to $E_v$ is smaller than that of $2\varepsilon_v$; thereby, it is called evaporative heating~\cite{Simula2014}. However, when the system is highly dissipative, $E_v$ can decrease quickly, even without decreasing $N_v$, thus lowering $\varepsilon_v$. Some theoretical studies raised a question regarding the fundamental possibility of $\varepsilon_v$ increasing in decaying QT~\cite{Numasato2010,Chesler2013}.

Atomic BECs with highly oblate geometry provide a suitable system for 2D QT~\cite{Neely2013,Kwon2014}, as the vortex line excitations are strongly suppressed along the tightly confining direction~\cite{Jackson2009,Rooney2011}. Many numerical studies have been performed using the Gross-Pitaevskii (GP) equation and have indicated that various turbulence regimes can exist in the system parameter space spanned by compressibility~\cite{Horng2009,Numasato2010}, dissipation~\cite{Bradley2012, Nowak2012,Kusumura2013,Reeves2013,Billam2014,Simula2014,Reeves2014,Billam2015,Stagg2015}, and trapping geometry~\cite{White2012,Groszek2016}. In previous experiments, vortex clustering was examined in a forced annular BEC~\cite{Neely2013}, and the thermal relaxation of turbulent BECs was investigated~\cite{Kwon2014}. However, full characterization of a turbulent BEC has never been achieved. Such a characterization requires measurements of not only the vortex positions but also their circulation directions. Vortex circulation signs might be determined by tracking the motions of individual vortices~\cite{Neely2010,Freilich2010} or by analyzing an interference fringe pattern with a stationary reference sample~\cite{Chevy2001,Inouye2001}, although this is experimentally challenging using a BEC with a complex vortex configuration. A new imaging technique was proposed in which a BEC is tilted before imaging so that each vortex core shows vortex sign-dependent deformation~\cite{Powis2014}.

In this study, we conduct spatially resolved Bragg spectroscopy to measure the full 2D vortex configuration of a turbulent BEC. Using this method, we examine the evolution of decaying 2D QT in a BEC at low temperatures, with our lowest being $\sim 0.5 T_c$, where $T_c$ is the critical temperature of the trapped sample. We observe the development of weak pair correlations between vortices and antivortices in the turbulent BEC, which corresponds to the vortex-dipole gas regime predicted for high dissipation. This work represents the first full experimental characterization of 2D QT in a BEC system and the results reported herein can be a valuable quantitative reference for theories of atomic superfluid turbulence.

\begin{figure}
\centering
\includegraphics[width=8.0cm]{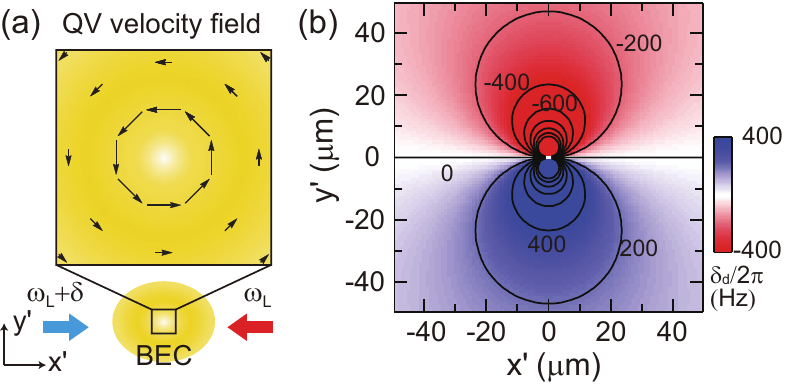}
\caption{Bragg scattering of a Bose-Einstein condensate (BEC) with a quantum vortex (QV). (a) The BEC is irradiated by two counterpropagating laser beams along the $x'$ direction with different frequencies of $\omega_L$ and $\omega_L+\delta$. (b) Bragg resonance frequency distribution around a singly charged QV. $\delta_d=\delta -\delta_0$, where $\delta_0$ is the resonance frequency for atoms at rest. As a result of the circulating velocity field, the resonance frequency is antisymmetric with respect to the Bragg scattering $x'$ axis. }
\label{fig2}
\end{figure}

\section{Results}

\subsection*{Vortex sign detection via Bragg scattering}

Our vortex sign detection method is based on the velocity sensitivity of Bragg scattering~\cite{Blakie2001}. Let us consider the situation where a BEC with a singly charged vortex is irradiated by a pair of counterpropagating laser beams along the $x'$ direction [Fig.~\ref{fig2}(a)]. A two-photon process, which imparts momentum $\vec{q}$ and energy $\varepsilon$ to an atom, occurs resonantly when $\varepsilon=q^2/2m+\vec{q} \cdot \vec{v}$, where $m$ and $\vec{v}$ are the atomic mass and velocity, respectively. Here, $\vec{q}=2\hbar k_L \hat{x'}$ and $\varepsilon=\hbar \delta$, where $k_L$ is the wavenumber of the two Bragg beams and $\delta$ is their frequency difference. For a positive vortex with counterclockwise circulation, the velocity field is given by $\vec{v}=\hbar/(mr^2) (\hat{z}\times \vec{r})$ with $\vec{r}$ being the position from the vortex core, and the resonance condition is given by $\delta_d=\delta-\delta_0 = -(2\hbar k_L /m)y'/r^2$, where $\delta_0=2 \hbar k_L^2/m$. Because of the Doppler effect, the scattering response is antisymmetric with respect to the Bragg beam axis~[Fig.~\ref{fig2}(b)]; thus, the vortex sign can be determined from the position of the scattered atoms relative to the vortex core. The use of Bragg scattering to measure a superfluid velocity field was demonstrated with a rotating BEC~\cite{Muniz2006}. In this work, we probe high-velocity regions near vortex cores to determine the circulation signs of individual vortices.

\begin{figure}[ht]
\centering
\includegraphics[width=8.4cm]{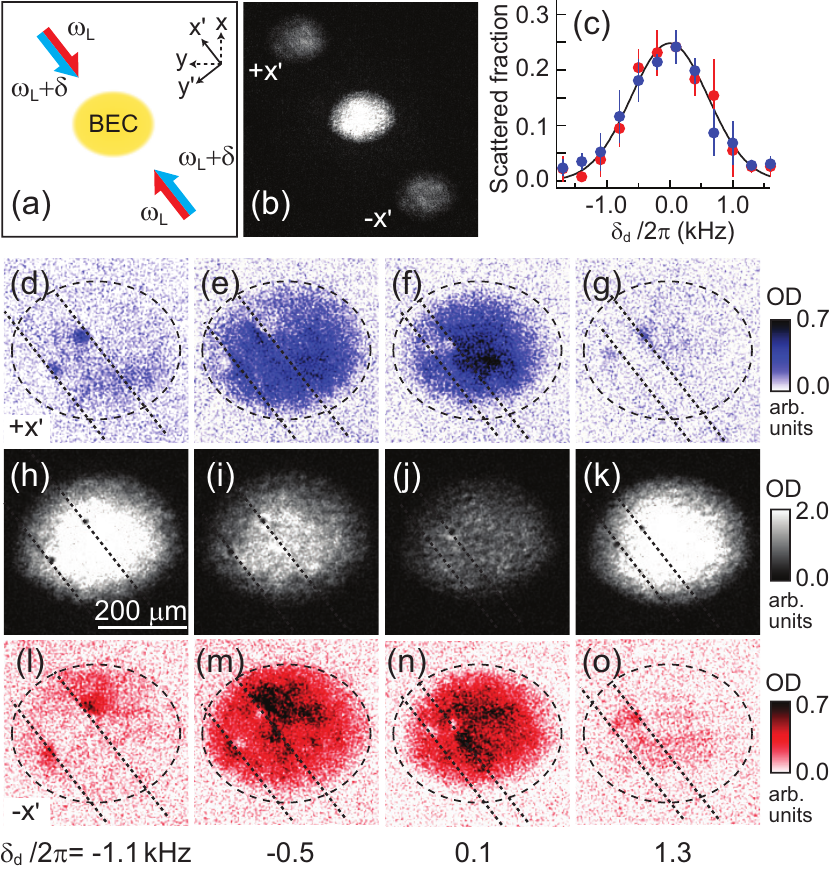}
\caption{Spatially resolved Bragg spectroscopy of a BEC. (a) Schematic of the experimental setup employing two pairs of counterpropagating Bragg beams. (b) Example of Bragg spectroscopy image for $\delta_d/2\pi=0.4$~kHz. Two atomic clouds are dispersed from the BEC in the $\pm x'$ directions. (c) Scattered-out atom number fractions measured for a stationary BEC as a function of $\delta_d$. The blue and red circles denote the atom number fractions of the $+x'$-and $-x'$-scattered atom clouds, respectively. The curved line is a Gaussian function of $A\exp[-(\delta-\delta_0)^2/(2\delta_w^2)]$ fit to the data, where $A=0.25$, $\delta_0/2\pi=99.2$~kHz, and $\delta_w/2\pi=615$ Hz. $\delta_w$ is accounted for by the finite pulse broadening. Bragg responses of a BEC having a vortex dipole for various frequencies $\delta_d$: (d)-(g) $+x'$-scattered atom clouds; (h)-(k) remaining condensates; and (l)-(o) $-x'$-scattered atom clouds. The dashed lines denote the boundary of the initial BEC, and the dotted lines indicate the Bragg beam lines that pass the vortex cores. The circulation signs of the vortices are known based on their trajectories in the trapped BEC, and the upper-right (lower-left) vortex has a positive (negative) circulation.}
\label{fig3}
\end{figure}

We conduct experiments using a BEC of $^{23}$Na atoms in the $|F=1,m_F=-1\rangle$ state in a pancake-shaped hybrid trap composed of optical and magnetic potentials. The trapping frequencies are $(\omega_x,\omega_y,\omega_z)=2\pi \times(4.3,3.5,350)$ Hz. For an atom number $N=4.0(3) \times 10^6$ and a condensate fraction of $80\%$, the Thomas-Fermi radii are $(R_x,R_y,R_z)=(155,190,1.9)~\mu$m. The condensate chemical potential is $\mu\approx h \times 510$~Hz and the healing length at peak density is $\xi=\hbar/\sqrt{2m \mu}\approx 0.6~\mu$m. The vortex dynamics is effectively 2D for $R_z/\xi\approx 3$~\cite{Jackson2009,Rooney2011}. Two pairs of Bragg beams are employed by retro-reflecting two laser beams with frequencies of $\omega_L$ and $\omega_L+\delta$, which are red-detuned by $\approx 1.7$~GHz from the $F=1$ to $F'=2$ transition [Fig.~\ref{fig3}(a)]. We apply a Bragg beam pulse for 600~$\mu$s after a short time of flight (TOF) of 300~$\mu$s which is initiated by releasing the trapping potential. During the short TOF, the condensate rapidly expands along the tightly confined $z$ direction to reduce the optical depth of the sample for the Bragg beams, but the modification of the transverse velocity field is negligible. After an additional TOF of $\tau=9$~ms, we take an absorption image of the sample~[Fig.~\ref{fig3}(b)]. Two atom clouds are scattered out from the condensate in both the $\pm x'$ directions. Since the displacement due to the initial atomic velocity is negligible, i.e., $v_{x'}\tau=|\delta_d|\tau /2k_L < 5~\mu$m for $|\delta_d|/2\pi<2$~kHz, the spatial distributions of the two scattered atom clouds reliably reveal the velocity regions that satisfy the Bragg scattering condition in the condensate.

We first apply spatially resolved Bragg spectroscopy to a BEC containing a vortex dipole, i.e., one positive and one negative vortex. A vortex dipole is generated by linearly sweeping the center region of the condensate using a repulsive Gaussian laser beam~\cite{Neely2010,Kwon2015,Kwon2016}, and after a period of 2~s, when the two vortices are well separated in the trapped condensate, we probe the sample using the Bragg beams. Figures~\ref{fig3}(d)-\ref{fig3}(o) display the density distributions of the condensate and the two scattered atom clouds for various values of $\delta_d$. The vortex positions are identified based on the density-depleted cores that appear in the condensate image. Note that the signs of each vortex are unambiguously known based on the vortex trajectories~\cite{Neely2010}; the upper-right (lower-left) vortex has a positive (negative) circulation. The scattering region becomes localized near the vortices with increasing $|\delta_d|$, indicating the existence of high-velocity regions in the proximity of the vortex cores. From the comparisons of the high-$|\delta_d|$ image data shown in Figs.~\ref{fig3}(d),~\ref{fig3}(g),~\ref{fig3}(l), and~\ref{fig3}(o), it is apparent that the position of the localized scattering region relative to the vortex core becomes inverted with respect to the Bragg beam axis when the vortex sign or the sign of $\delta_d$ is changed or when the scattering direction is reversed. This is consistent with the aforementioned antisymmetric response of the vortex state to the Bragg scattering. Furthermore, we find that the density profiles of the scattered atom clouds near the vortex cores are quantitatively accounted for by a theoretical estimation including the spectral broadening of the Bragg scattering (see Supplementary Information).

\begin{figure}[b]
\centering
\includegraphics[width=8.0cm]{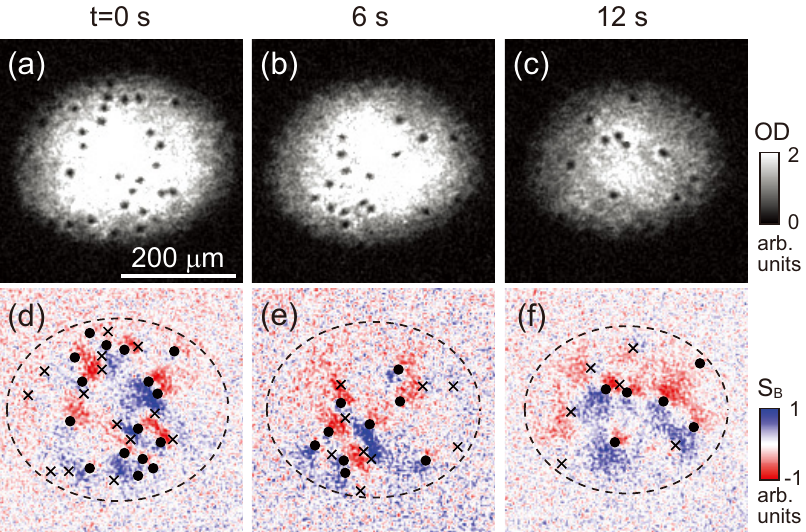}
\caption{Determination of the vortex configuration of a turbulent BEC. (a)-(c) TOF images of BECs at various hold times $t$ and (d)-(f) the corresponding Bragg signals $S_B(x',y')=n_+-n_-$, where $n_\pm$ are the density distributions of the $\pm x'$-scattered atom clouds. The vortex positions are identified based on the density-depleted holes that appear in the BEC images, and their circulation signs are determined based on $S_B$ around the vortex cores (see the text). The circles and crosses denote positive and negative vortices, respectively.}
\label{fig4}
\end{figure}

\begin{figure*}
\centering
\includegraphics[width=17.0cm]{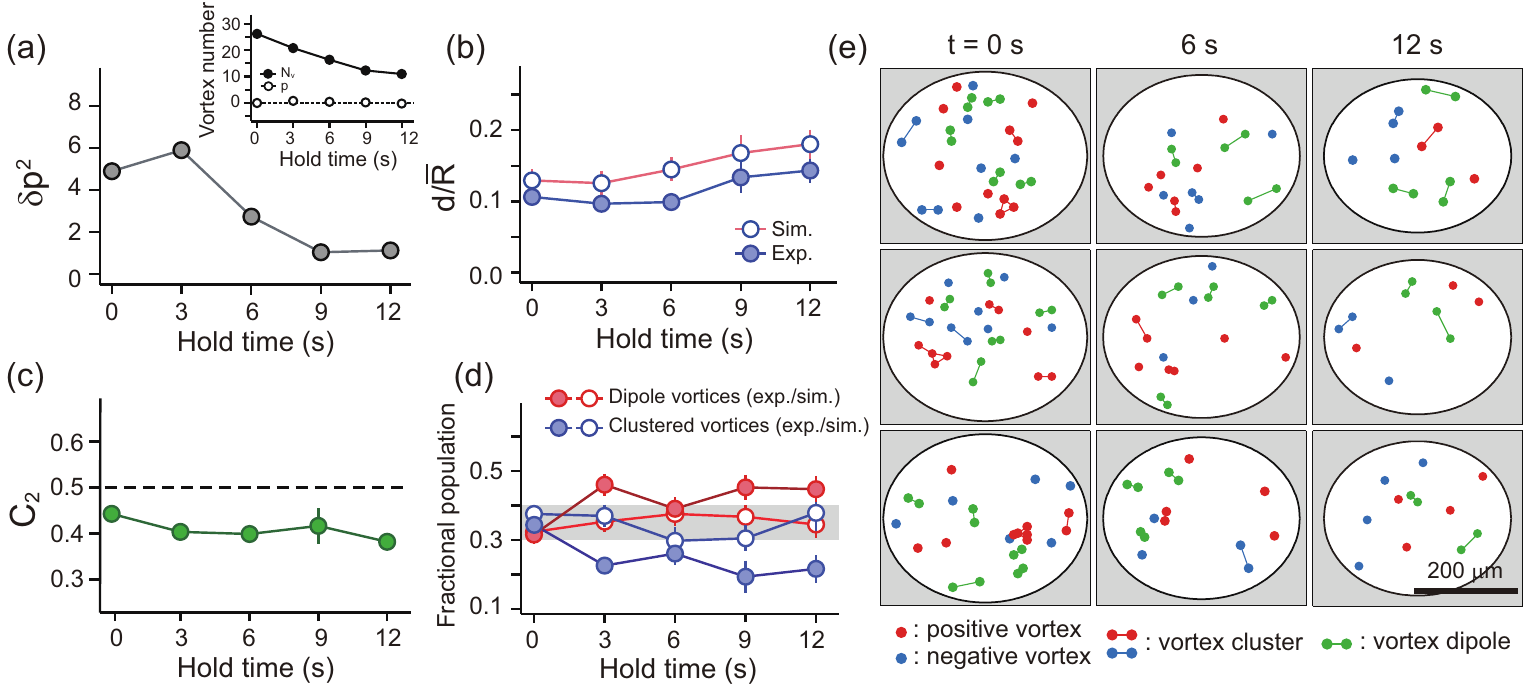}
\caption{Characterization of decaying 2D quantum turbulence. Various properties of the turbulent BEC were measured based on its vortex configuration as a function of the hold time: (a) vortex polarization variance, $\delta p^2$, where the inset shows $N_v=N_{+}+N_{-}$ and $p=N_{+}-N_{-}$, and $N_{\pm}$ is the number of positive (negative) vortices; (b) mean vortex dipole moment $d$ divided by the condensate radius $\bar{R}=\sqrt{R_x R_y}$; (c) second-order vortex sign correlation function $C_2$~\cite{White2012}; and (d) fractional populations of dipole vortices and clustered vortices~\cite{Billam2014,Billam2015}. Examples of the vortex configuration data including the vortex classification results are displayed in (e) for various hold times $t$. Each data point in (a)-(d) was obtained from fifteen to twenty measurements of the same experiment, and each point's error bar indicates the standard error of the mean of the measurements. The open circles in (b) and (d) show the simulation results calculated using twenty vortex configurations randomly sampled for the same $N_{\pm}$. The gray region in (d) indicates the range of the simulation results. }
\label{fig5}
\end{figure*}

\subsection*{Probing 2D quantum turbulence}

Next, we apply the Bragg scattering method to probe the vortex configuration of a turbulent BEC containing a large number of vortices. Turbulence is generated by stirring the condensate using a repulsive laser beam (see the Methods section). The initial vortex number is $N_v\approx 26$ and the mean intervortex distance is $l_v\sim \bar{R}/\sqrt{N_v}\approx 34~\mu$m, where $\bar{R}=\sqrt{R_x R_y}$. We set $\delta_d/2\pi=-1.1$~kHz, which was observed in the previous experiment to yield a localized scattering signal peaking at $r\approx 13~\mu$m from a vortex core [Fig.~\ref{fig3}(d)]. A higher $|\delta_d|$ generates a more localized signal but the signal-to-noise ratio is poor.

To facilitate the vortex sign determination, we construct a Bragg signal $S_B(x',y')\equiv n_{+}-n_{-}$, where $n_{\pm}(x',y')$ are the density distributions of the $\pm x'$-scattered atom clouds, which are translated to the condensate reference frame [Figs.~\ref{fig4}(d)-\ref{fig4}(f)]. Because $n_{+}$ and $n_{-}$ are complementary to each other due to the local mirror symmetry along the Bragg beam line [Figs.~\ref{fig3}(d) and \ref{fig3}(l)], $S_B$ contains vortex-sign information. In $S_B$, the vortex sign is manifested as the sign of the signal derivative along the $y'$ direction at the vortex position $(x'_i,y'_i)$, i.e., a positive (negative) vortex appears for a positive (negative) value of $\partial S_B /\partial y'$. We determine the sign of $\partial S_B /\partial y'$ by evaluating $\int_{-a}^{a}  \mathrm{sgn}(y) S_B(x'_i,y'_i+y) dy$ with $a=13~\mu$m. When many vortices are located in close proximity to each other, the Bragg signal around some vortices might be weak, and it would be necessary to scrutinize the overall vortex configuration to assign the vortex signs (see Supplementary Information). In particular, in the case of a small vortex dipole for which the surrounding velocity field is almost canceled, the scattering signal is absent, and the vortex signs must then be determined based on the crescent shape of their merged density-depleted cores~\cite{Kwon2014}. Thus, $S_B$ and the condensate density distribution provide sufficient information to determine the full vortex configuration of the turbulent BEC.

\subsection*{Vortex-antivortex pairing}

The complete determination of the vortex configuration enables us to characterize the evolution of the turbulent state of the BEC (Fig.~\ref{fig5}). The BEC relaxes as the vortex number $N_v$ decreases, and the vortex half-life time is $\approx10$~s [Fig.~\ref{fig5}(a) inset]. As expected from the vortex sign symmetry, the vortex polarization $p=N_+ - N_-$ maintains a zero-mean value, where $N_{\pm}$ are the numbers of positive and negative vortices, respectively. Interestingly, we observe that the polarization variance, $\delta p^2$, decreases during the evolution [Fig.~\ref{fig5}(a)]. If the vortex decay is a vortex-sign-independent process, $\delta p^2$ would increase as $\delta p^2(t)=\delta p^2(0)+[N_v(0)-N_v(t)]$, similar to diffusion by a random walk. The reduction of $\delta p^2$ indicates that a polarized turbulent state is forced to decay into a balanced state and it also suggests that vortex-antivortex pair annihilation is the dominant vortex decay mechanism in a turbulent BEC.

We measure the mean vortex dipole moment of the BEC, $d=|\frac{1}{N_v}\sum^{N_v}_{i=1} s_i \vec{r}_i|$, where $s_i=\pm 1$ is the sign of the $i$th vortex and $\vec{r}_i$ is its position with respect to the condensate center [Fig.~\ref{fig5}(b)]. In our experiment, $d\approx0.1\bar{R}$ throughout the decay evolution. The measured value of $d$ is found to be slightly smaller than the mean value obtained for the random vortex distributions sampled for the same $N_\pm$, thus excluding the formation of the Onsager vortex state. We note that in the vortex-dipole gas regime, $d$ can be smaller than the mean dipole moment of random distributions for finite $N_v$.

To examine the vortex pair correlations, we evaluate the second-order vortex sign correlation function $C_2=\frac{1}{2 N_v}\sum^{N_v}_{i=1}\sum^{2}_{j=1}c_{ij}$, where $c_{ij}=1(0)$ if the $i$th vortex and its $j$th nearest neighbor have the same (different) sign~\cite{White2012}. A truly random configuration yields $C_2=0.5$ and like-sign vortex clusters and vortex dipoles are reflected as increases and decreases of $C_2$, respectively. Our experimental data show that $C_2\approx 0.4$~[Fig.~\ref{fig5}(d)], indicating that it is more probable to have neighboring vortices with opposite signs.

We perform a further analysis of the measured vortex configurations by applying the vortex classification algorithm introduced by Billam {\it et al.}~\cite{Billam2014,Billam2015}: two vortices are assigned as a dipole if they are the nearest neighbors to each other and have opposite signs; a group of same-sign vortices as a cluster if they are closer to each other than to any other opposite-sign vortex; and the remaining vortices as free vortices [Fig.~\ref{fig5}(e)-~\ref{fig5}(g)]. In recent numerical studies, it was shown that the fractional populations of dipole vortices, clustered vortices, and free vortices according to this classification scheme provide a unique representation of the 2D QT states, suggested as vortex thermometry~\cite{Groszek2017}. We measure the vortex numbers, $N_d$, $N_c$, and $N_f$, of dipoles, clusters, and free vortices, respectively, where $N_d+N_c+N_f=N_v$. The initial turbulence state shows $N_d\approx N_c \approx N_f$, which is a characteristic of the random vortex configuration~\cite{Groszek2017}, and it is observed that as the decay evolution proceeds, the fractional population of dipole vortices increases to $\frac{N_d}{N_v}\approx 0.45$, whereas that of clustered vortices decreases to $\frac{N_c}{N_v}\approx 0.2$. This observation corroborates the vortex-antivortex pairing in the turbulent BEC.

\section{Discussion}

All the results of our vortex configuration analysis demonstrate that vortex-antivortex pair correlations develop in a turbulent BEC under our experimental conditions. It was anticipated that the characteristics of 2D QT in atomic BECs evolve into the vortex-dipole gas regime as the system's dissipation becomes stronger~\cite{Reeves2014,Simula2014,Billam2015}, but since the quantitative understanding of dissipation in finite-temperature vortex dynamics is still incomplete~\cite{Rooney2010}, there is no theoretical prediction regarding the critical temperature at which the emergence of the Onsager vortex state can be observed~\cite{Kim2016}. Thermal damping is typically modeled using a few parameters~\cite{Berloff2014}, but it might be questionable whether the damping effects in various vortex dynamics can be fully captured by the parameters. Our experimental results provide quantitative information regarding the dissipative vortex dynamics in 2D QT. For future reference, our main finding is summarized as follows: a highly oblate turbulent BEC with $N_v\approx 20$ and $\bar{R}/\xi\approx 290$ evolves at $T/T_c\approx 0.5$ to a state with $C_2\approx 0.4$, $N_d/N_v\approx 0.45$, and $N_c/N_v\approx 0.2$.

This work can be extended to investigate various 2D QT regimes by changing the system parameters. Although it is highly desirable to reach a low dissipation regime by lowering the sample temperature, this was difficult to achieve in our experiment because the sample was heated during the turbulence generation process. It has been noted that utilizing a steep-wall trap instead of a harmonic trap provides a beneficial condition for the formation of Onsager vortex state~\cite{Groszek2016}. Additionally, it might be conceivable to prepare an Onsager vortex state by merging two oppositely rotating BECs and to investigate its relaxation through the vortex-dipole gas regime at high temperatures.

In summary, we have demonstrated the Bragg scattering method for detecting the quantum vortex circulation sign and have successfully applied it to probe decaying 2D QT in a trapped BEC. Various properties of the turbulent BEC were measured based on its vortex configuration and the development of vortex-antivortex pairing was observed in our experiment at finite temperatures. We expect that the Bragg scattering method presented here will enable a direct experimental study of various 2D QT regimes in the atomic BEC system.

\section*{Acknowledgements}

This work was supported by IBS-R009-D1 and the National Research Foundation of Korea (Grant No. 2013-H1A8A1003984).


\begin{thebibliography}{10}
\expandafter\ifx\csname url\endcsname\relax
  \def\url#1{\texttt{#1}}\fi
\expandafter\ifx\csname urlprefix\endcsname\relax\def\urlprefix{URL }\fi
\expandafter\ifx\csname doiprefix\endcsname\relax\def\doiprefix{DOI }\fi
\providecommand{\bibinfo}[2]{#2}
\providecommand{\eprint}[2][]{\url{#2}}

\bibitem{Skrbek2012}
\bibinfo{author}{Skrbek, L.} \& \bibinfo{author}{Sreenivasan, K.~R.}
\newblock \bibinfo{journal}{\bibinfo{title}{Developed quantum turbulence and
  its decay}}.
\newblock {\emph{{Physics of Fluids}}}
  \textbf{\bibinfo{volume}{24}}, \bibinfo{pages}{011301}
  (\bibinfo{year}{2012}).

\bibitem{Tsubota2013}
\bibinfo{author}{Tsubota, M.}, \bibinfo{author}{Kobayashi, M.} \&
  \bibinfo{author}{Takeuchi, H.}
\newblock \bibinfo{journal}{\bibinfo{title}{Quantum hydrodynamics}}.
\newblock {\emph{{Phys. Rep.}}}
  \textbf{\bibinfo{volume}{522}}, \bibinfo{pages}{191--238}
  (\bibinfo{year}{2013}).

\bibitem{Henn2009}
\bibinfo{author}{Henn, E. A.~L.}, \bibinfo{author}{Seman, J.~A.},
  \bibinfo{author}{Roati, G.}, \bibinfo{author}{Magalhaes, K. M.~F.} \&
  \bibinfo{author}{Bagnato, V.~S.}
\newblock \bibinfo{journal}{\bibinfo{title}{{Emergence of Turbulence in an
  Oscillating Bose-Einstein Condensate}}}.
\newblock {\emph{{Phys. Rev. Lett.}}}
  \textbf{\bibinfo{volume}{103}}, \bibinfo{pages}{045301}
  (\bibinfo{year}{2009}).

\bibitem{Neely2013}
\bibinfo{author}{Neely, T.~W.} \emph{et~al.}
\newblock \bibinfo{journal}{\bibinfo{title}{Characteristics of two-dimensional
  quantum turbulence in a compressible superfluid}}.
\newblock {\emph{{Phys. Rev. Lett.}}}
  \textbf{\bibinfo{volume}{111}}, \bibinfo{pages}{235301}
  (\bibinfo{year}{2013}).

\bibitem{Kwon2014}
\bibinfo{author}{Kwon, W.~J.}, \bibinfo{author}{Moon, G.},
  \bibinfo{author}{Choi, J.}, \bibinfo{author}{Seo, S.~W.} \&
  \bibinfo{author}{Shin, Y.}
\newblock \bibinfo{journal}{\bibinfo{title}{Relaxation of superfluid turbulence
  in highly oblate Bose-Einstein condensates}}.
\newblock {\emph{{Phys. Rev. A}}} \textbf{\bibinfo{volume}{90}},
  \bibinfo{pages}{063627} (\bibinfo{year}{2014}).

\bibitem{Navon2016}
\bibinfo{author}{Navon, N.}, \bibinfo{author}{Gaunt, A.~L.},
  \bibinfo{author}{Smith, R.~P.} \& \bibinfo{author}{Hadzibabic, Z.}
\newblock \bibinfo{journal}{\bibinfo{title}{Emergence of a turbulent cascade in
  a quantum gas}}.
\newblock {\emph{Nature}} \textbf{\bibinfo{volume}{539}},
  \bibinfo{pages}{72--75} (\bibinfo{year}{2016}).

\bibitem{Krichnan1980}
\bibinfo{author}{Kraichnan, R.~H.} \& \bibinfo{author}{Montgomery, D.}
\newblock \bibinfo{journal}{\bibinfo{title}{{Two-dimensional turbulence}}}.
\newblock {\emph{{Rep. Prog. Phys.}}}
  \textbf{\bibinfo{volume}{43}}, \bibinfo{pages}{547--619}
  (\bibinfo{year}{1980}).

\bibitem{Horng2009}
\bibinfo{author}{Horng, T.-L.}, \bibinfo{author}{Hsueh, C.-H.},
  \bibinfo{author}{Su, S.-W.}, \bibinfo{author}{Kao, Y.-M.} \&
  \bibinfo{author}{Gou, S.-C.}
\newblock \bibinfo{journal}{\bibinfo{title}{Two-dimensional quantum turbulence
  in a nonuniform bose-einstein condensate}}.
\newblock {\emph{{Phys. Rev. A}}} \textbf{\bibinfo{volume}{80}},
  \bibinfo{pages}{023618} (\bibinfo{year}{2009}).

\bibitem{Numasato2010}
\bibinfo{author}{Numasato, R.}, \bibinfo{author}{Tsubota, M.} \&
  \bibinfo{author}{L'vov, V.~S.}
\newblock \bibinfo{journal}{\bibinfo{title}{Direct energy cascade in
  two-dimensional compressible quantum turbulence}}.
\newblock {\emph{{Phys. Rev. A}}} \textbf{\bibinfo{volume}{81}},
  \bibinfo{pages}{063630} (\bibinfo{year}{2010}).

\bibitem{White2012}
\bibinfo{author}{White, A.~C.}, \bibinfo{author}{Barenghi, C.~F.} \&
  \bibinfo{author}{Proukakis, N.~P.}
\newblock \bibinfo{journal}{\bibinfo{title}{Creation and characterization of
  vortex clusters in atomic Bose-Einstein condensates}}.
\newblock {\emph{{Phys. Rev. A}}} \textbf{\bibinfo{volume}{86}},
  \bibinfo{pages}{013635} (\bibinfo{year}{2012}).

\bibitem{Bradley2012}
\bibinfo{author}{Bradley, A.~S.} \& \bibinfo{author}{Anderson, B.~P.}
\newblock \bibinfo{journal}{\bibinfo{title}{Energy spectra of vortex
  distributions in two-dimensional quantum turbulence}}.
\newblock {\emph{{Phys. Rev. X}}} \textbf{\bibinfo{volume}{2}},
  \bibinfo{pages}{041001} (\bibinfo{year}{2012}).

\bibitem{Nowak2012}
\bibinfo{author}{Nowak, B.}, \bibinfo{author}{Schole, J.},
  \bibinfo{author}{Sexty, D.} \& \bibinfo{author}{Gasenzer, T.}
\newblock \bibinfo{journal}{\bibinfo{title}{Nonthermal fixed points, vortex
  statistics, and superfluid turbulence in an ultracold Bose gas}}.
\newblock {\emph{{Phys. Rev. A}}} \textbf{\bibinfo{volume}{85}},
  \bibinfo{pages}{043627} (\bibinfo{year}{2012}).

\bibitem{Kusumura2013}
\bibinfo{author}{Kusumura, T.}, \bibinfo{author}{Takeuchi, H.} \&
  \bibinfo{author}{Tsubota, M.}
\newblock \bibinfo{journal}{\bibinfo{title}{{Energy spectrum of the superfluid
  velocity made by quantized vortices in two-dimensional quantum turbulence}}}.
\newblock {\emph{{Journal of Low Temperature Physics}}}
  \textbf{\bibinfo{volume}{171}}, \bibinfo{pages}{563--570}
  (\bibinfo{year}{2013}).

\bibitem{Reeves2013}
\bibinfo{author}{Reeves, M.~T.}, \bibinfo{author}{Billam, T.~P.},
  \bibinfo{author}{Anderson, B.~P.} \& \bibinfo{author}{Bradley, A.~S.}
\newblock \bibinfo{journal}{\bibinfo{title}{Inverse energy cascade in forced
  two-dimensional quantum turbulence}}.
\newblock {\emph{{Phys. Rev. Lett.}}}
  \textbf{\bibinfo{volume}{110}}, \bibinfo{pages}{104501}
  (\bibinfo{year}{2013}).

\bibitem{Chesler2013}
\bibinfo{author}{Chesler, P.~M.}, \bibinfo{author}{Liu, H.} \&
  \bibinfo{author}{Adams, A.}
\newblock \bibinfo{journal}{\bibinfo{title}{Holographic vortex liquids and
  superfluid turbulence}}.
\newblock {\emph{Science}} \textbf{\bibinfo{volume}{341}},
  \bibinfo{pages}{368--372} (\bibinfo{year}{2013}).

\bibitem{Simula2014}
\bibinfo{author}{Simula, T.}, \bibinfo{author}{Davis, M.~J.} \&
  \bibinfo{author}{Helmerson, K.}
\newblock \bibinfo{journal}{\bibinfo{title}{Emergence of order from turbulence
  in an isolated planar superfluid}}.
\newblock {\emph{{Phys. Rev. Lett.}}}
  \textbf{\bibinfo{volume}{113}}, \bibinfo{pages}{165302}
  (\bibinfo{year}{2014}).

\bibitem{Reeves2014}
\bibinfo{author}{Reeves, M.~T.}, \bibinfo{author}{Billam, T.~P.},
  \bibinfo{author}{Anderson, B.~P.} \& \bibinfo{author}{Bradley, A.~S.}
\newblock \bibinfo{journal}{\bibinfo{title}{Signatures of coherent vortex
  structures in a disordered two-dimensional quantum fluid}}.
\newblock {\emph{{Phys. Rev. A}}} \textbf{\bibinfo{volume}{89}},
  \bibinfo{pages}{053631} (\bibinfo{year}{2014}).

\bibitem{Billam2014}
\bibinfo{author}{Billam, T.~P.}, \bibinfo{author}{Reeves, M.~T.},
  \bibinfo{author}{Anderson, B.~P.} \& \bibinfo{author}{Bradley, A.~S.}
\newblock \bibinfo{journal}{\bibinfo{title}{Onsager-Kraichnan condensation in
  decaying two-dimensional quantum turbulence}}.
\newblock {\emph{{Phys. Rev. Lett.}}}
  \textbf{\bibinfo{volume}{112}}, \bibinfo{pages}{145301}
  (\bibinfo{year}{2014}).

\bibitem{Billam2015}
\bibinfo{author}{Billam, T.~P.}, \bibinfo{author}{Reeves, M.~T.} \&
  \bibinfo{author}{Bradley, A.~S.}
\newblock \bibinfo{journal}{\bibinfo{title}{Spectral energy transport in
  two-dimensional quantum vortex dynamics}}.
\newblock {\emph{{Phys. Rev. A}}} \textbf{\bibinfo{volume}{91}},
  \bibinfo{pages}{023615} (\bibinfo{year}{2015}).

\bibitem{Stagg2015}
\bibinfo{author}{Stagg, G.~W.}, \bibinfo{author}{Allen, A.~J.},
  \bibinfo{author}{Parker, N.~G.} \& \bibinfo{author}{Barenghi, C.~F.}
\newblock \bibinfo{journal}{\bibinfo{title}{Generation and decay of
  two-dimensional quantum turbulence in a trapped Bose-Einstein condensate}}.
\newblock {\emph{{Phys. Rev. A}}} \textbf{\bibinfo{volume}{91}},
  \bibinfo{pages}{013612} (\bibinfo{year}{2015}).

\bibitem{Du2015}
\bibinfo{author}{Du, Y.}, \bibinfo{author}{Niu, C.},
  \bibinfo{author}{Tian, Y.} \& \bibinfo{author}{Zhang, H.}
\newblock \bibinfo{journal}{\bibinfo{title}{Holographic thermal relaxation in superfluid turbulence}}.
\newblock {\emph{{J. High Energ. Phys.}}} \textbf{\bibinfo{volume}{12}},
  \bibinfo{pages}{018} (\bibinfo{year}{2015}).
  
\bibitem{Skaugen2016}
\bibinfo{author}{Sakugen, A.} \& \bibinfo{author}{Angheluta, L.} 
\newblock \bibinfo{journal}{\bibinfo{title}{Vortex clustering and universal scaling laws in two-dimensional quantum turbulence}}.
\newblock{\emph{{Phys. Rev. E}}} \textbf{\bibinfo{volume}{93}}, \bibinfo{pages}{032106} (\bibinfo{year}{2016}).

\bibitem{Nazarenko14}
\bibinfo{author}{Nazarenko, S.}, \bibinfo{author}{Onorato, M.} \&
  \bibinfo{author}{Proment, D.}
\newblock \bibinfo{journal}{\bibinfo{title}{Bose-Einstein condensation and
  Berezinskii-Kosterlitz-Thouless transition in the two-dimensional nonlinear
  schr\"odinger model}}.
\newblock {\emph{{Phys. Rev. A}}} \textbf{\bibinfo{volume}{90}},
  \bibinfo{pages}{013624} (\bibinfo{year}{2014}).

\bibitem{Onsager1949}
\bibinfo{author}{Onsager, L.}
\newblock \bibinfo{journal}{\bibinfo{title}{{Statistical hydrodynamics}}}.
\newblock {\emph{{Il Nuovo Cimento Series 9}}}
  \textbf{\bibinfo{volume}{6}}, \bibinfo{pages}{279--287}
  (\bibinfo{year}{1949}).

\bibitem{Eyink2006}
\bibinfo{author}{Eyink, G.~L.} \& \bibinfo{author}{Sreenivasan, K.~R.}
\newblock \bibinfo{journal}{\bibinfo{title}{Onsager and the theory of
  hydrodynamic turbulence}}.
\newblock {\emph{{Rev. Mod. Phys.}}}
  \textbf{\bibinfo{volume}{78}}, \bibinfo{pages}{87--135}
  (\bibinfo{year}{2006}).

\bibitem{Yatsuyanagi2005}
\bibinfo{author}{Yatsuyanagi, Y.} \emph{et~al.}
\newblock \bibinfo{journal}{\bibinfo{title}{Dynamics of two-sign point vortices
  in positive and negative temperature states}}.
\newblock {\emph{{Phys. Rev. Lett.}}}
  \textbf{\bibinfo{volume}{94}}, \bibinfo{pages}{054502}
  (\bibinfo{year}{2005}).

\bibitem{Demircan1996}
\bibinfo{author}{Demircan, E.}, \bibinfo{author}{Ao, P.} \&
  \bibinfo{author}{Niu, Q.}
\newblock \bibinfo{journal}{\bibinfo{title}{Vortex dynamics in superfluids: Cyclotron-type motion}}.
\newblock {\emph{{Phys. Rev. B}}} \textbf{\bibinfo{volume}{54}},
  \bibinfo{pages}{10027--10034} (\bibinfo{year}{1996}).

\bibitem{Kobayashi2006}
\bibinfo{author}{Kobayashi, M.} \& \bibinfo{author}{Tsubota, M.}
\newblock \bibinfo{journal}{\bibinfo{title}{Thermal dissipation in quantum
  turbulence}}.
\newblock {\emph{{Phys. Rev. Lett.}}}
  \textbf{\bibinfo{volume}{97}}, \bibinfo{pages}{145301}
  (\bibinfo{year}{2006}).

\bibitem{Berloff2007}
\bibinfo{author}{Berloff, N.~G.} \& \bibinfo{author}{Youd, A.~J.}
\newblock \bibinfo{journal}{\bibinfo{title}{Dissipative dynamics of superfluid
  vortices at nonzero temperatures}}.
\newblock {\emph{{Phys. Rev. Lett.}}}
  \textbf{\bibinfo{volume}{99}}, \bibinfo{pages}{145301}
  (\bibinfo{year}{2007}).

\bibitem{Moon2015}
\bibinfo{author}{Moon, G.}, \bibinfo{author}{Kwon, W.~J.},
  \bibinfo{author}{Lee, H.} \& \bibinfo{author}{Shin, Y.}
\newblock \bibinfo{journal}{\bibinfo{title}{Thermal friction on quantum vortices in a Bose-Einstein condensate}}.
\newblock {\emph{{Phys. Rev. A}}} \textbf{\bibinfo{volume}{92}},
  \bibinfo{pages}{051601} (\bibinfo{year}{2015}).

\bibitem{Jackson2009}
\bibinfo{author}{Jackson, B.}, \bibinfo{author}{Proukakis, N.~P.},
  \bibinfo{author}{Barenghi, C.~F.} \& \bibinfo{author}{Zaremba, E.}
\newblock \bibinfo{journal}{\bibinfo{title}{Finite-temperature vortex dynamics
  in Bose-Einstein condensates}}.
\newblock {\emph{{Phys. Rev. A}}} \textbf{\bibinfo{volume}{79}},
  \bibinfo{pages}{053615} (\bibinfo{year}{2009}).

\bibitem{Rooney2011}
\bibinfo{author}{Rooney, S.~J.}, \bibinfo{author}{Blakie, P.~B.},
  \bibinfo{author}{Anderson, B.~P.} \& \bibinfo{author}{Bradley, A.~S.}
\newblock \bibinfo{journal}{\bibinfo{title}{Suppression of kelvon-induced decay of quantized vortices in oblate Bose-Einstein condensates}}.
\newblock {\emph{{Phys. Rev. A}}} \textbf{\bibinfo{volume}{84}},
  \bibinfo{pages}{023637} (\bibinfo{year}{2011}).

\bibitem{Groszek2016}
\bibinfo{author}{Groszek, A.~J.}, \bibinfo{author}{Simula, T.~P.},
  \bibinfo{author}{Paganin, D.~M.} \& \bibinfo{author}{Helmerson, K.}
\newblock \bibinfo{journal}{\bibinfo{title}{Onsager vortex formation in Bose-Einstein condensates in two-dimensional power-law traps}}.
\newblock {\emph{{Phys. Rev. A}}} \textbf{\bibinfo{volume}{93}},
  \bibinfo{pages}{043614} (\bibinfo{year}{2016}).

\bibitem{Neely2010}
\bibinfo{author}{Neely, T.~W.}, \bibinfo{author}{Samson, E.~C.},
  \bibinfo{author}{Bradley, A.~S.}, \bibinfo{author}{Davis, M.~J.} \&
  \bibinfo{author}{Anderson, B.~P.}
\newblock \bibinfo{journal}{\bibinfo{title}{Observation of vortex dipoles in an
  oblate Bose-Einstein condensate}}.
\newblock {\emph{{Phys. Rev. Lett.}}}
  \textbf{\bibinfo{volume}{104}}, \bibinfo{pages}{160401}
  (\bibinfo{year}{2010}).

\bibitem{Freilich2010}
\bibinfo{author}{Freilich, D.~V.}, \bibinfo{author}{Bianchi, D.~M.},
  \bibinfo{author}{Kaufman, A.~M.}, \bibinfo{author}{Langin, T.~K.} \&
  \bibinfo{author}{Hall, D.~S.}
\newblock \bibinfo{journal}{\bibinfo{title}{{Real-time dynamics of single vortex lines and vortex dipoles in a Bose-Einstein condensate.}}}
\newblock {\emph{Science}} \textbf{\bibinfo{volume}{329}},
  \bibinfo{pages}{1182-1185} (\bibinfo{year}{2010}).

\bibitem{Chevy2001}
\bibinfo{author}{Chevy, F.}, \bibinfo{author}{Madison, K.~W.},
  \bibinfo{author}{Bretin, V.} \& \bibinfo{author}{Dalibard, J.}
\newblock \bibinfo{journal}{\bibinfo{title}{Interferometric detection of a
  single vortex in a dilute Bose-Einstein condensate}}.
\newblock {\emph{{Phys. Rev. A}}} \textbf{\bibinfo{volume}{64}},
  \bibinfo{pages}{031601} (\bibinfo{year}{2001}).

\bibitem{Inouye2001}
\bibinfo{author}{Inouye, S.} \emph{et~al.}
\newblock \bibinfo{journal}{\bibinfo{title}{Observation of vortex phase singularities in Bose-Einstein condensates}}.
\newblock {\emph{{Phys. Rev. Lett.}}}
  \textbf{\bibinfo{volume}{87}}, \bibinfo{pages}{080402}
  (\bibinfo{year}{2001}).


\bibitem{Powis2014}
\bibinfo{author}{Powis, A.~T.}, \bibinfo{author}{Sammut, S.~J.} \&
  \bibinfo{author}{Simula, T.~P.}
\newblock \bibinfo{journal}{\bibinfo{title}{Vortex gyroscope imaging of planar superfluids}}.
\newblock {\emph{{Phys. Rev. Lett.}}}
  \textbf{\bibinfo{volume}{113}}, \bibinfo{pages}{165303}
  (\bibinfo{year}{2014}).

\bibitem{Blakie2001}
\bibinfo{author}{Blakie, P.~B.} \& \bibinfo{author}{Ballagh, R.~J.}
\newblock \bibinfo{journal}{\bibinfo{title}{Spatially selective bragg scattering: A signature for vortices in Bose-Einstein condensates}}.
\newblock {\emph{{Phys. Rev. Lett.}}}
  \textbf{\bibinfo{volume}{86}}, \bibinfo{pages}{3930--3933}
  (\bibinfo{year}{2001}).

\bibitem{Muniz2006}
\bibinfo{author}{Muniz, S.~R.}, \bibinfo{author}{Naik, D.~S.} \&
  \bibinfo{author}{Raman, C.}
\newblock \bibinfo{journal}{\bibinfo{title}{Bragg spectroscopy of vortex lattices in Bose-Einstein condensates}}.
\newblock {\emph{{Phys. Rev. A}}} \textbf{\bibinfo{volume}{73}},
  \bibinfo{pages}{041605} (\bibinfo{year}{2006}).

\bibitem{Kwon2015}
\bibinfo{author}{Kwon, W.~J.}, \bibinfo{author}{Seo, S.~W.} \&
  \bibinfo{author}{Shin, Y.}
\newblock \bibinfo{journal}{\bibinfo{title}{Periodic shedding of vortex dipoles from a moving penetrable obstacle in a Bose-Einstein condensate}}.
\newblock {\emph{{Phys. Rev. A}}} \textbf{\bibinfo{volume}{92}},
  \bibinfo{pages}{033613} (\bibinfo{year}{2015}).

\bibitem{Kwon2016}
\bibinfo{author}{Kwon, W.~J.}, \bibinfo{author}{Kim, J.~H.},
  \bibinfo{author}{Seo, S.~W.} \& \bibinfo{author}{Shin, Y.}
\newblock \bibinfo{journal}{\bibinfo{title}{Observation of von K\'arm\'an  vortex street in an atomic superfluid gas}}.
\newblock {\emph{{Phys. Rev. Lett.}}}
  \textbf{\bibinfo{volume}{117}}, \bibinfo{pages}{245301}
  (\bibinfo{year}{2016}).



\bibitem{Groszek2017}
\bibinfo{author}{Groszek, A.~J.}, \bibinfo{author}{Davis, M.~J.},
  \bibinfo{author}{Paganin, D.~M.}, \bibinfo{author}{Helmerson, K.} \&
  \bibinfo{author}{Simula, T.~P.}
\newblock \bibinfo{journal}{\bibinfo{title}{{Vortex thermometry for turbulent two-dimensional fluids}}}.
\newblock {\emph{{arXiv:1702.05229}}}.

\bibitem{Rooney2010}
\bibinfo{author}{Rooney, S.~J.}, \bibinfo{author}{Bradley, A.~S.} \&
  \bibinfo{author}{Blakie, P.~B.}
\newblock \bibinfo{journal}{\bibinfo{title}{Decay of a quantum vortex: Test of  nonequilibrium theories for warm Bose-Einstein condensates}}.
\newblock {\emph{{Phys. Rev. A}}} \textbf{\bibinfo{volume}{81}},
  \bibinfo{pages}{023630} (\bibinfo{year}{2010}).

\bibitem{Kim2016}
\bibinfo{author}{Kim, J.~H.}, \bibinfo{author}{Kwon, W.~J.} \&
  \bibinfo{author}{Shin, Y.}
\newblock \bibinfo{journal}{\bibinfo{title}{Role of thermal friction in relaxation of turbulent Bose-Einstein condensates}}.
\newblock {\emph{{Phys. Rev. A}}} \textbf{\bibinfo{volume}{94}},
  \bibinfo{pages}{033612} (\bibinfo{year}{2016}).

\bibitem{Berloff2014}
\bibinfo{author}{Berloff, N.~G.}, \bibinfo{author}{Brachet, M.} \&
  \bibinfo{author}{Proukakis, N.~P.}
\newblock \bibinfo{journal}{\bibinfo{title}{Modeling quantum fluid dynamics at nonzero temperatures}}.
\newblock {\emph{{Proceedings of the National Academy of
  Sciences}}} \textbf{\bibinfo{volume}{111}}, \bibinfo{pages}{4675--4682}
  (\bibinfo{year}{2014}).

\end{thebibliography}

\section*{SUPPLEMENTAL INFORMATION}

\subsection*{Vortex state preparation}

We generated quantum vortices by stirring the center region of a BEC using a focused repulsive Gaussian laser beam as demonstrated in previous experiments~\cite{Neely2010,Kwon2014,Kwon2015,Kwon2016}. The $1/e^2$ beam width was $\sigma \approx 10 \mu$m and the potential barrier height was $V \approx h\times 8$ kHz. When we stirred the condensate, the radial trapping frequencies were $\omega_{x,y}/2\pi=7.5$ Hz. A tighter trap is helpful for minimizing the dipole motion of the condensate, which might be induced by the stirring. After the vortex generation, the radial trapping potential was adiabatically ramped down within 2 s to the condition of the main experiment. To generate a vortex dipole, we linearly swept the condensate by translating the laser beam in the $-y$ direction over $\approx 100~\mu$m with a velocity of $v=0.98$ mm/s $\approx0.25c_s$, where $c_s$ is the speed of sound. For generating turbulence, we stirred the condensate in a sinusoidal manner with an amplitude of $40~\mu$m at 15~Hz for 200~ms.

\subsection*{Spatial distribution of scattered atoms near a vortex}

At a position of $\vec{r}$ from a vortex with sign $s_v=\pm 1$, the atom velocity is $\vec{v}(\vec{r})= s_v \hbar/ (m r^2) (\hat{z}\times \vec{r} )$, and for resonant Bragg scattering the frequency difference $\delta_v(\vec{r})$ of the two laser bemas  is given by $\hbar \delta_v (\vec{r})=q^2/(2m)+\vec{q}\cdot \vec{v}(\vec{r})$. From $\hbar \delta_0 = q^2/(2m)$ and $\vec{q}=\pm \sqrt{2m\hbar \delta_0} \hat{x}'$, we have $\delta_v(\vec{r})=\delta_0 \mp s_v \sqrt{2 \hbar \delta_0/m} (y'/r^2).$
Taking into account the spectral broadening of the Bragg scattering, which is measured with a stationary BEC in Fig.~2(c), we estimate the spatial distribution of the scattered atoms as
\begin{eqnarray}
n_\pm(\vec{r})&=&n_c(\vec{r}) A \exp \Big[ -\frac{\big(\delta-\delta_v(\vec{r})\big)^2}{\delta_w^2} \Big] \nonumber \\ 
&=& n_0 \frac{A r^2}{r^2+r_c^2} \exp \Big[ -\frac{\big( \delta_d \pm s_v \sqrt{\frac{2 \hbar \delta_0}{m}} (y'/r^2) \big)^2}{\delta_w^2} \Big], 
\end{eqnarray}
where $n_c(\vec{r})=n_0 r^2/(r^2+r_c^2)$ is the initial density distribution of the condensate. In our experiment, $A=0.25$ and $\delta_w/2\pi=615$~Hz. We set $r_c=\sqrt{2}\xi'$, where $\xi'$ is the healing length of the condensate after 300~$\mu$s TOF. In Fig.~7, we compare the experimental data with the theoretical estimation, and find them in good quantitative agreement, including possible blurring effects due to the finite imaging resolution and the expansion during the TOF.

\begin{figure}
\includegraphics[width=7.6cm]{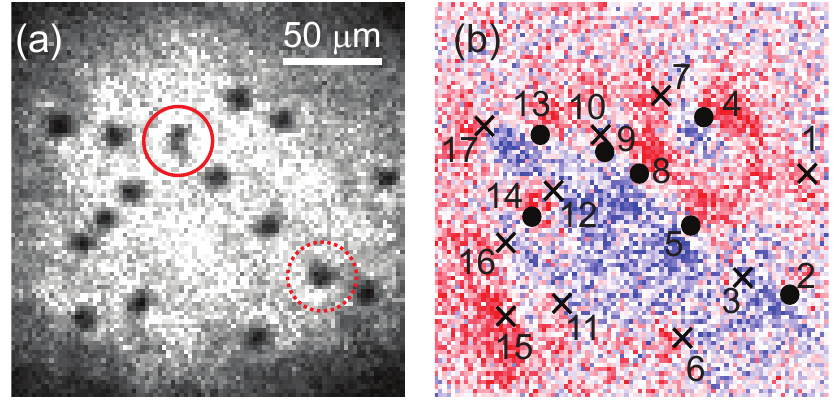}
\caption{Vortex sign determination. (a) Sample image of a turbulent BEC and (b) the corresponding Bragg signal $S_B(x',y')$. For most of the vortices, their signs are unambiguously determined from the signs of the quantity $\int_{-a}^{a} \mathrm{sgn}(y') S_B(x'_i,y'_i+y') dy'$, as described in the main text. However, in the cases of vortex 3 (red dashed circle), and that of vortex 9 and 10 (red solid circle), the Bragg signal around the vortices is too weak to immediately determine their signs. To vortex 3, we assign a negative sign by examining the vortex configuration in its surrounding. If vortex 3 is positive, the Bragg signal around vortex 2 and 3 should have a similar pattern to that around vortex 5 and 8. Also, we note that the faint blue-colored (positive $S_B$) region between vortex 2 and 3 is not compatible with vortex 3 being positive. Vortex 9 and 10 form a tight vortex dipole, showing a crescent-shaped density depleted core region~\cite{Kwon2014}. The curvature of the crescent shape indicates that the vortices are moving to the left direction and thus, the signs of vortex 9 and 10 are positive and negative, respectively.}
\label{FigS2}
\end{figure}

\begin{figure*}
\includegraphics[width=14cm]{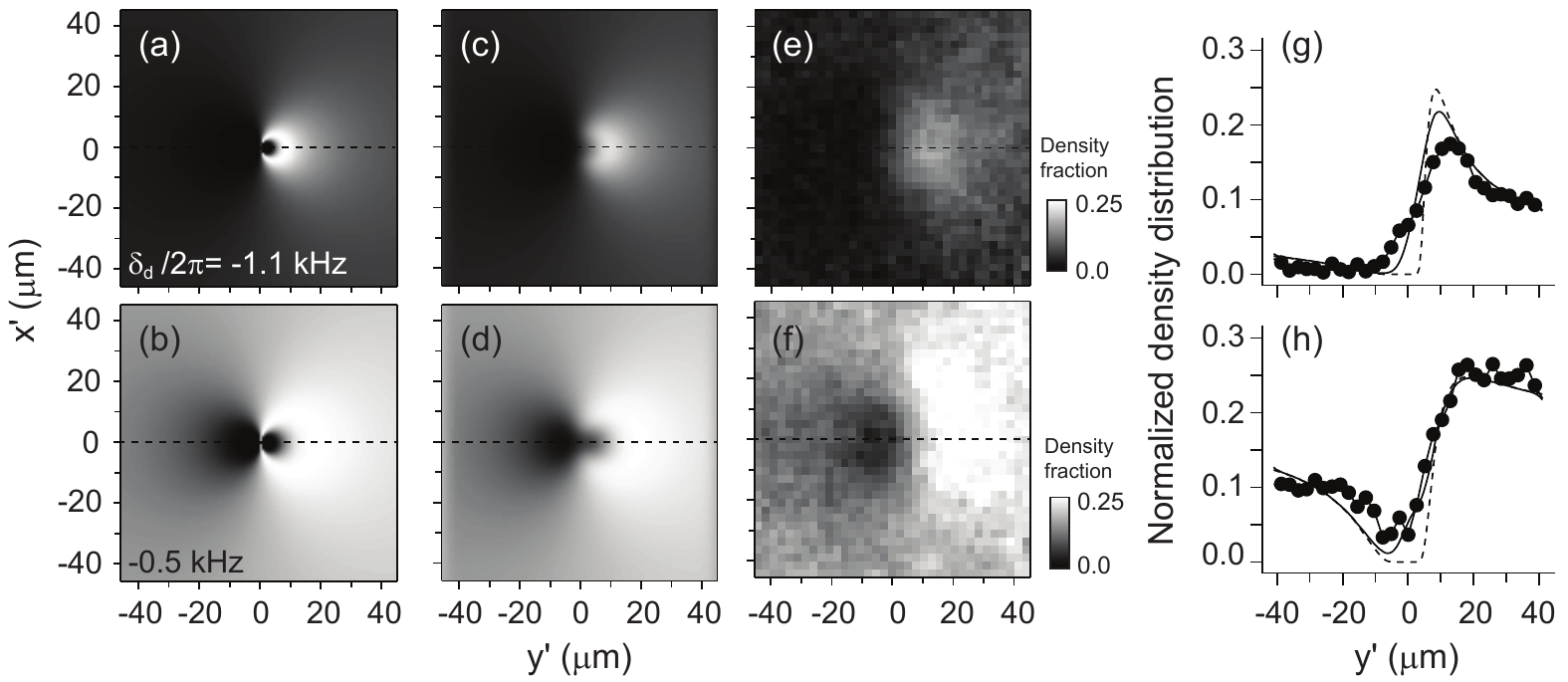}
\caption{Scattered atom density distribution. $n_{+}(x',y')/n_0$ calculated from Eq.~(1) for $s_v=-1$, and $\delta_d/2\pi=-1.1$~kHz (a) and $-0.5$~kHz (b). (c)-(d) Gaussian-blurred images of (a) and (b), respectively. The blurring width was set to be 10~$\mu$m, which is comparable to the imaging resolution. Experimental data of $n_+(x',y')/n_0$ for $\delta_d/2\pi=-1.1$~kHz (e) and $-0.5$~kHz (f), obtained by averaging over ten image data of the vortex region cropped from scattered atom cloud images. To minimize the effect from other vortices, we used only image data where vortices are separated by over $100~\mu$m. $n_0$ was determined from a Thomas-Fermi profile fit to the total atom density distribution of the initial condensate. (g)-(h) Density proflies along the $y'$-axis indicated by the horizontal dashed lines in (a)-(f): the experimental data (solid circles), and the theoretical calculation results with and without blurring (solid and dashed lines, respectivley).}
\label{FigS1}
\end{figure*}

\subsection*{Incompressible kinetic energy spectrum}

The incompressible kinetic energy spectrum $E(k)$ as a function of wavenumber $k$ is the essential measure for a turbulence state and the energy flow direction and large structure formation can be reflected in its evolution~\cite{Billam2015}. Bradley {\it et al.}~\cite{Bradley2012} presented the analytic expression of $E(k)$ for a given vortex configuration $\{\vec{r}_i, s_i\}$ as
\begin{equation}
E(k)=N_vE_0F(sk\xi)[1+\frac{2}{N_v}\sum_{i=1}^{N_v-1}\sum_{j=i+1}^{N_v}s_is_jJ_0(k|\vec{r_i}-\vec{r_j}|)], 
\end{equation}
where $E_0= s h^2\tilde{n}\xi/2\pi m$ with $s=1.25$ and $\tilde{n}$ being the column density, and $F(z)= z/4[I_1(z/2)K_0(z/2)-I_0(z/2)K_1(z/2)]^2$. $J_0$ is the zeroth order Bessel function, and $I_i$ and $K_i$ are the $i$~th-order modified Bessel functions of the first and second kinds, respectively. Figure~8 shows $E(k)$ obtained for our experimental data using Eq.~(2). $E(k)$ exhibits $k^{-3}$ scaling for $k>1/\xi$, which corresponds to the vortex core structure~\cite{Bradley2012}; it smoothly changes to $k^{-1}$ scaling in the range of $2\pi/\bar{R}<k<1/\xi$, which is the characteristic behavior for a random vortex configuration due to the $1/r$ velocity field around a vortex~\cite{Bradley2012}. In our measurements, there is no indication of the inertial region with $k^{-5/3}$ scaling.

\begin{figure}
	\includegraphics[width=7.8cm]{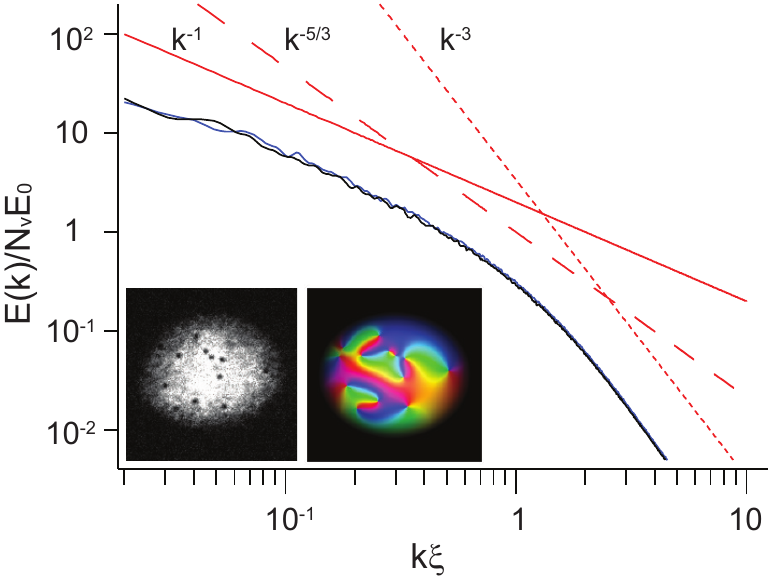}
	\caption{ Incompressible kinetic energy spectrum $E(k)$. $E(k)$ is calculated from the measured vortex configuration using Eq. (2). The blue and black lines are the averaged spectra of the experimental data for $t=3$ s and 9~s, respectively. The insets display an image the condensate at $t=9$~s (left) and its phase distribution reconstructed from the vortex configuration information (right).}
	\label{fig_s3}
\end{figure}

\end{document}